\documentclass[preprint,12pt]{elsarticle}
\usepackage{gensymb,amssymb,amsmath}
\usepackage{color,xcolor}

\begin{document}

\begin{frontmatter}

\title{Social Influence and Consensus Building: Introducing a q-Voter Model with Weighted Influence}

\author[label1]{Pratik Mullick}
\author[label2]{Parongama Sen}

\affiliation[label1]{organization={Department of Operations Research and Business Intelligence, Wrocław University of Science and Technology},
            addressline={Wyb. Wyspiańskiego 27}, 
            city={50-370},
            postcode={Wrocław}, 
            state={Lower Silesia},
            country={Poland}}
            
\affiliation[label2]{organization={Department of Physics, University of Calcutta},
            addressline={92 Acharya Prafulla Chandra Road}, 
            city={700009},
            postcode={Kolkata}, 
            state={West Bengal},
            country={India}}

\begin{abstract}

We investigate a dynamical model of opinion formation in which an individual's opinion is influenced by interactions with a group of other agents.
We introduce a bias towards one of the opinions in a manner not considered earlier to the best of our knowledge.
When the bias is neutral, the model is reduced to a mean-field voter model. We analyze the behavior and steady states of the system, identifying three distinct regimes based on the bias level: one favoring negative opinions, one favoring positive opinions, and a neutral case. In large systems, the equilibrium properties become independent of the size of the group, indicating that only the bias influences the final outcome. However, for small groups, the time to reach equilibrium depends on the size of the group. Our results show that even a small initial bias leads to a consensus where all agents eventually share the same opinion when the bias is not neutral. The system exhibits universal behavior, with critical slowing down occurring near the neutral bias point, marking it as a critical dynamical threshold. The time required to reach consensus scales logarithmically when the bias is non-neutral and linearly when it is neutral. Although short-term dynamics depends on group size for small groups, long-term behavior is governed solely by the bias.  
\end{abstract}



\begin{keyword}
social dynamics \sep consensus formation \sep influential power \sep opinion dynamics \sep voter model \sep mean field theory \sep Monte Carlo simulation 


\end{keyword}

\end{frontmatter}

\section{Introduction}\label{intro}

Alice works in a company as a data scientist. The company policy for decision making is very employee-friendly. The authorities value the opinion of each of their employees. Before taking any major decision, they hold a meeting, propose the idea to their employees and accept it, if and only if all the employees agree to it, i.e., if there is consensus. However, in most of the cases the opinions among the employees are mixed, some agree with taking the decision, some do not. In this case, the company gives more time to their employees to rethink and arrange another meeting after a while to take the decision or to reject it. Within this time, the employees discuss among themselves about the decision, some stick to their original opinion, whereas some change it when sufficiently convinced by oppositely opinionated colleagues. Alice is given the responsibility to conduct a study to prescribe how a positive consensus could be achieved. She has to estimate how much the convincing ability or influential power of the agents with positive opinion should be such that they manage to flip the decisions of all the agents with negative opinion.

The formation of consensus \cite{Susskind1999, Rayens2000,Innes2004,Caillaud2007,Goodfriend2007,Lee2007,Gillard2016,Dong2017,Urena2019,Lu2021,Foroughi2023,han2023large,liu2023optimizing,Qin2023,Liu2024,Sievers2024} is a collaborative process that aims to achieve general agreement within a group. It involves open communication, mutual respect, and active participation from all members. Unlike majority voting \cite{Oliveira1992,Krapivsky2003dynamics,Chen2005,Lima2006,Lima2012,Vilela2012}, consensus building seeks to address and integrate diverse perspectives, often through negotiation and compromise, to reach a decision that everyone can support, or at least accept. This approach is commonly used in settings where collective decision-making \cite{bouzarour2015bipolar,Lu2021,han2023large,Qin2023,liu2023optimizing,Liu2023grey,Liu2024,Peng2024} is crucial, such as in community planning \cite{Foroughi2023,Qin2023}, organizational environment \cite{Lee2007, Dong2017}, social networks \cite{Urena2019,Lu2021}, and policy development \cite{Rayens2000,Goodfriend2007,Gillard2016}. Collective decision-making in a social group is vital as it harnesses diverse perspectives, leading to more informed and well-rounded decisions. It fosters a sense of ownership and commitment among group members, enhancing cooperation and cohesion. This inclusive process also promotes transparency and accountability, reducing conflicts and increasing the likelihood of successful implementation. Moreover, collective decision-making leverages the collective intelligence of the group, often resulting in more innovative and effective solutions to complex problems.

Modelling opinion dynamics can be a valuable method for studying consensus building \cite{Fortunato2005,Yang2009,Dolfin2015,Dong2017,Bauso2018,Liu2021,Hassani2022} or collective decision-making \cite{Montes2011,Moussaid2013,Urena2019review,Li2021,li2021multi,Dong2022,han2023large,Liu2023grey,Peng2024,pei2024conflict}. 
 By analyzing the processes of social influence \cite{Moussaid2013,Quattrociocchi2014,Javarone2014,Colaiori2015}, peer interactions \cite{Urbig2008opinion,Kindler2013peer,Liu2021model}, and information dissemination \cite{Das2014modeling,Li2020analysis,Xu2020dynamic,Geng2023online}, researchers can understand how consensus emerges from diverse viewpoints. In models for opinion formation, the main emphasis is how peer interaction takes place.  
 Other factors such as the role of opinion leaders \cite{Zhao2016bounded,Dong2017,Glass2021social,Weng2023integrating}, the impact of network structures \cite{Gabbay2007effects,Dolfin2015,Han2020clusters,Wang2020new}, and the effects of external factors \cite{Sirbu2017opinion,Jkedrzejewski2018impact,Li2020effect}, such as mass media \cite{civitarese2021external,muslim2024mass} are also taken into account. This approach helps identify the conditions under which consensus is more likely to be achieved, the mechanisms driving opinion convergence, and the potential barriers to collective decision-making.

 The basic models of opinion dynamics \cite{Castellano2009rmp} provide frameworks for understanding how individual opinions evolve and aggregate within a social group, contributing to consensus building and collective decision-making. 
 The DeGroot Model \cite{Degroot1974reaching} involves individuals updating their opinions based on a weighted average of their neighbors' opinions, demonstrating how repeated interactions lead to convergence. Bounded Confidence Models \cite{Deffuant2000mixing,Hegselmann2002opninion,wang2024neural} allow interactions only within a certain opinion range, illustrating how clusters form and consensus or polarization occurs. 
 One of the earliest models, namely the  Voter Model \cite{Holley1975ergodic,Sood2005,Castellano2009qvm}, has individuals randomly adopting a neighbor's opinion, showing how majority opinions emerge over time. The Sznajd Model \cite{Sznajd2000} emphasizes social validation, where individuals adopt the opinion of a pair of agreeing neighbors, highlighting local consensus effects. 
 The Hegselmann-Krause Model \cite{Hegselmann2002opninion,Hegselmann2005opninion,Hegselmann2006opninion}, similar to bounded confidence models, but uses the average opinion of all neighbors within a confidence interval to update opinions, effectively analyzing the impact of openness on consensus formation. The Majority Rule Model \cite{Oliveira1992,Krapivsky2003dynamics,Chen2005,Lima2006,Lima2012,Vilela2012} is another fundamental framework where each individual adopts the opinion of the majority within their local group, illustrating how consensus can emerge from local majority interactions. In all these models an important quantity studied is the exit probability $E(x)$ which denotes the probability that the system would reach a positive consensus after starting from $x$ fraction of agents with positive opinion.

 The research area of opinion dynamics has inherent connections with statistical physics, where the challenge lies in understanding a social phenomena using models and methods from theoretical physics. The Ising Model, pivotal in statistical physics for describing ferromagnetism, analogously represents individuals as binary spins on a lattice, which can align based on local interactions and external influences. This model's principles are mirrored in opinion dynamics \cite{Grabowski2006ising,Biswas2009model,Sen2014sociophysics,Dai2022ising}, where individuals' opinions are influenced by their neighbors, similar to spin alignment. Models like voter model and Sznajd model could be seen as variants of the Ising model, borrowing concepts from statistical physics to explain how local interactions propagate and lead to macroscopic patterns in opinion distribution. These models can be studied on various network structures, including lattices, random graphs, small-world networks, and scale-free networks, each representing different types of social connectivity and interaction patterns. The complete graph is particularly significant as it represents an idealized scenario where every individual can directly interact with every other individual.

 In this context, mean-field theory of statistical physics offers a simplified analytical approach in understanding complex systems by averaging the effects of all interactions \cite{Barabasi1999mean,Kardar2007statistical,Aoki2014nonequilibrium}. This theory assumes that each individual's influence is equally distributed across the entire network, allowing the system to be described by a single average field rather than accounting for detailed interactions. In opinion dynamics, mean-field theory is used to approximate the behavior of models on large, well-mixed populations \cite{Castellano2009qvm,Fennell2021generalized,LIPIECKI2022,Maciek2022}. It effectively captures the overall trend of opinion alignment and consensus formation by considering the average effect of social influence, rather than the detailed network structure. This approach is particularly useful for studying systems on complete graphs, where each individual interacts equally with every other individual, thereby justifying the assumption of uniform influence and simplifying the analysis of consensus dynamics.


\subsection{Literature review: \textit{q} voter model and beyond}
The original voter model \cite{Holley1975ergodic} with binary opinions, where an agent randomly selects one of the neighbour's opinion, exhibits linear behaviour in exit probability as $E(x)=x$. 
A nonlinear voter model was presented by Castellano et. al. \cite{Castellano2009qvm}, where a selected agent interacts with $q$ other neighbours. When the $q$-panel is unanimous, the agent selects their opinion - a situation known as conformity, otherwise the opinion is flipped with a probability. This model is also known as the $q$-voter model in literature. The dynamics are different from that of majority rule model \cite{galam2008sociophysics,krapivsky2021divergence}, where a $q$-panel is selected, and all the members belonging to the panel take the majority opinion within them.
 In \cite{czaplicka2022biased}, the agents are initially segmented into two groups - biased and unbiased - that remain fixed throughout the dynamics. The unbiased agents exhibit original voter dynamics, and the biased agents flip their opinions with probabilities that depend on their original own opinion and the opinion of their neighbour.
 Another interesting variant of the voter model is the noisy voter model where spontaneous flippings of the opinions are allowed \cite{GRANOVSKY1995}.
 
 There are a plethora of journal articles that have studied variants of the non-linear or $q$-voter model \cite{Castellano2009qvm}, varying the original definition in one way or the other. Here we provide a non-exhaustive overview of such studies that have been done on various network structures. In \cite{byrka2016difficulty}, a model was studied on complete graph to understand social diffusion of innovation where the randomly chosen agent becomes either independent (not influenced by her $q$ neighbours) or becomes conformist with the complementary probability. In \cite{mobilia2015nonlinear}, a $q$ voter model was studied on a complete graph with some `zealots' in the system, who are inflexible with their opinions and do not change state under any conditions; with the susceptible agents maintaining conformist behaviour with a unanimous $q$-panel. Nyczka et. al. studied three different models in \cite{nyczka2012phase}, each with conformist, anti-conformist (focal agent takes the opinion opposite to the one in unanimous $q$-panel) and independent agents respectively. A variant of $q$ voter model was studied in \cite{javarone2015conformism} where the agents are either conformist with a probability, or anti-conformist with the complementary probability; without considering any of them to be independent. A conformist $q$ voter model was also studied on a one dimensional lattice with periodic boundary conditions \cite{timpanaro2014exit}, where a consecutively indexed $q$-panel was chosen randomly. In case of unanimity in this $q$ neighbourhood, either both the adjacent agents, or one of the 2 adjacent agents conforms to their opinion.
 
 Variants of the $q$ voter model were also studied on multiplex networks \cite{gradowski2020pair}, duplex clique \cite{chmiel2015phase}, Erdos-Renyi graphs \cite{vieira2020pair} and scale free networks \cite{vieira2020pair}. A typical variant called the threshold $q$ voter model was studied in \cite{vieira2018threshold} on a complete graph, where unanimity within the $q$-panel for a minimum number $q_0$ of agents ($0\leq q_0 \leq q$) is sufficient to influence the focal agent, at the same time keeping the possibility of its independence. This threshold $q$-voter model was later studied on random networks in \cite{vieira2020pair}. Muslim et. al. \cite{muslim2024mass} studied a $q$-voter model where in case of non-unanimity in the $q$-panel the focal agent chooses an opinion expressed by the mass media with a probability. A similar model was studied in \cite{civitarese2021external}, where the independent agent might become skeptical of its own opinion, triggered by an unreliable external field in social processes, quite similar to how mass media influences the decision making in our society.

\subsection{Our contribution: A new voter model with individual influential power}
 In this paper, we propose a $q$-voter model where the independent behaviour of the focal agent is absent in general. Rather the $q$-panel, even in case of non-unanimity, influences the opinion of the agent under consideration, depending on the ratio of the two opinions in this panel. Precisely, we consider that the  agents have an influential power depending on their current opinion (positive or negative), with which they  influence the opinion of the chosen agent.   The weighted influential powers of both types of opinionated agents decide the opinion of the focal agent. 

Rest of the paper is organised as follows. In the next section we define the dynamics of our model. In section \ref{results} we discuss our main results, where in section \ref{mfa} we use mean-field theory to find analytical solutions of our model. Followed by results from Monte Carlo simulations \ref{mcs} and their comparisons with analytical expressions. Finally in section \ref{conclu} we make some concluding remarks.

\section{Model Description and features studied}\label{model}

 In this model we consider a population of $L$ agents with binary opinions on a complete graph. The opinions are either positive or negative. Let us consider a situation when an agent $A$ is interacting with $q \geq 2$ other agents. If this $q$-panel is unanimous, $A$ takes their opinion, a situation known as conformity. When the $q$-panel is not unanimous, we consider $p$, the \textit{influential power} of agents with positive opinion. Naturally the influential power of agents with negative opinion is $1-p$. In the $q$-panel, let there be $n$ agents with positive opinion. These $n$ agents with influential power $p$ each, would try to convince $A$ with a total influential power of $np$. On the other hand, $q-n$ agents with negative opinion each with influential power $1-p$, would have a total influential power of $(q-n)(1-p)$.

\begin{figure}[h!]
    \centering
    \includegraphics[width=0.8\linewidth]{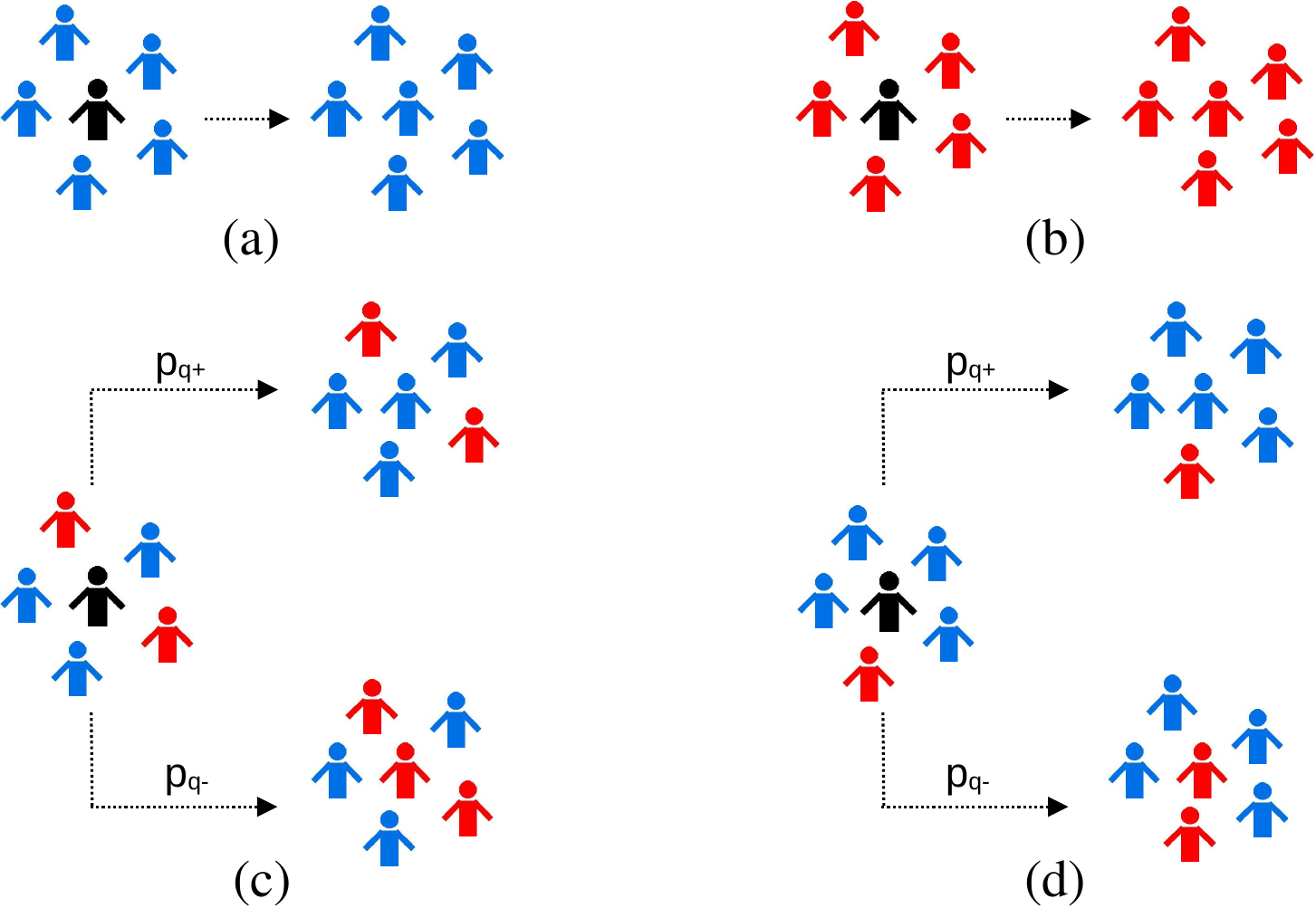}
    \caption{Schematic diagram showing the dynamics of our agent-based modelling. Here we show a typical case when a selected agent (shown in black) interacts with randomly chosen $5$ other agents, i.e., $q=5$. Case (a) shows conformity, when the $q$-panel of selected agent is comprised of all agents with positive opinion (shown in blue). Similarly, case (b) also represents conformity, the $q$-panel has only agents with negative opinion (shown in red). In case (c), the selected agent is surrounded by 3 agents with positive opinion and 2 agents with negative opinion, and in case (d) it is surrounded by 4 agents with positive opinion and 1 negative agent. In both these cases, the selected agent chooses positive opinion with probability $p_{q+}$ and negative opinion with probability $p_{q-}.$ The expressions for $p_{q+}$ and $p_{q-}$ are given by Eqs. (\ref{pq_plus}) and (\ref{pq_minus}) respectively.} 
    \label{model_schematic}
\end{figure}

 Since $A$ could take either the positive opinion or the negative opinion, the total probability of taking these two opinions at an elementary update should be 1. So if $p_{q+}$ and $p_{q-}$ denote the probabilities that $A$ could take the positive opinion and the negative opinion respectively, then \begin{align}
    p_{q+}&=\frac{np}{p(2n-q)+(q-n)}\label{pq_plus}\\[0.3cm]
    p_{q-}&=\frac{(q-n)(1-p)}{p(2n-q)+(q-n)}.\label{pq_minus}
\end{align}
$p_{q+}$ and $p_{q-}$ could be seen as the weighted average of influential powers of $n$ agents with positive opinion and $q-n$ agents with negative opinion. The term $np+(q-n)(1-p)=p(2n-q)+(q-n)$ is the normalisation factor. The dynamics of the model are schematically shown in Figure \ref{model_schematic}. 
It can be easily checked from Eqs. (\ref{pq_plus}) and (\ref{pq_minus}) that if $n=0$ or $q$, there is conformity. For $n=q/2$ (assuming $q$ is even), $p_{q+} = p$, clearly indicating the bias even when agents with the two opinions are present in equal number.  

For $q=2$ and $p=1/2$ the dynamics are identical to that of a  $q$-voter model with $q=2$ \cite{Castellano2009qvm}. 
In fact for  general values of $q$ and  $p=1/2$,  when all the agents are equally influential, $p_{q+} = n/q$, such that the model is equivalent to  a mean field voter model with $q$ neighbours in which an agent picks up any random neighbour and changes her opinion accordingly. When  the $q$ plaquette is unanimous, all the neighbours have the same state so the choice is unique. In this model, the $q$ neighbours vary, so in a fully connected network, one can say that at each time, out of the whole population, the agent interacts with $q$ other agents chosen randomly, which is closer to reality. Hence for $p=1/2$ one can expect a conservation as present in the voter model leading to a linear behavior of the exit probability.


Let $f_+(t)$ be the fraction of agents with positive opinion in the whole system at any time $t$.  The initial fraction  of agents with positive opinion is denoted by  $x=f_+(0)$ throughout this paper. 
Obviously, $f_{-}(t)$, the fraction of agents with negative opinions is equal to $1-f_+(t)$. 
We use mean field theory to obtain the dynamical equations governing $f_+(t)$
for different values of $q$ and solve the equations. We also make a fixed point analysis and a linear stability analysis for small $q$ values as well as for $q \to \infty$.

Using the Monte Carlo simulations, we study $f_+(t)$ as a function of time and compare with the mean field results. 
In general, we find that the final state is a consensus state with  all the opinions becoming either positive or negative. Hence one can also study the exit probability $E(x)$ here. An estimate of time scales has been made from both the mean field theory and numerical simulations. We present also the results for  the time taken to reach the consensus state as a function
of the system size in the simulations. 
 All quantitative studies are performed by varying the two independent parameters $q$ and $p$.

\section{Results and Discussion}\label{results}

\subsection{Mean field approach}\label{mfa}

 We use the Mean-field theory to find analytical results of our model, especially to have a theoretical understanding of how the underlying dynamics depend on $p$ and $q$. 

 For small values of $q$, it is possible to find the rate equations considering all possible cases $n = 0,1,2,\dots,q$ and we first present the results for the two cases $q=2$ and $q=3$ where all possibilities have been taken care of.
  For larger $q$ values, it is difficult to consider all the cases individually and 
  it is convenient to replace $n$ by an average value  $n=qf_+(t)$. This corresponds to
  the assumption  that the  fraction
of agents with positive opinions in the whole population is also maintained within the $q$-panel.
This assumption is the key point of the analytical calculations for $q > 3$.

It should be mentioned here that $f_+(t), f_-(t)$ are ensemble averaged quantities in the mean field theory. Also, we would drop the argument for brevity henceforth.

\subsubsection{$q=2$:}
\label{secq=2} In this case the value of $n$ is zero, 1 or 2. The fraction of positive opinions increases if an agent with a negative opinion interacts with a panel with $n=2$ (this panel occurs with probability $f_+^2$) and decreases
when an agent with positive opinion interacts with a panel  with $n=0$ (occurring with probability $1-f_{-}^2$). For $n=1$, Eqs. (\ref{pq_plus}) and (\ref{pq_minus}) give $p_{2+} = p$ and 
$p_{2-} = 1-p$ such that considering all possible cases one gets
\begin{equation}
\frac{df_+}{dt}=f_+^2(1-f_+)-(1-f_+)^2f_+ + 2f_+ (1-f_+)\big[p(1-f_+) -(1-p)f_+\big],
\end{equation} where the first two terms represent the cases $n=2$ ans $n=0$ respectively, and the last term represents the case $n=1$ which can occur in two ways.
The above equation simplifies to 
\begin{equation}
\frac{df_+}{dt}=f_+(1-f_+)(2p-1),
\label{q=2}
\end{equation}
which shows that there are only two fixed points for any $p \neq 0.5$ at $f_+ = 0,1$. For 
$p=0.5$ any value of $f_+$ is a fixed point, i.e., the opinions do not change at all. 

Let us now consider a model, which we call the  binary model with stochastic biased flipping or BMSBF.
Here, only pairwise interactions are allowed and if an agent with positive opinion interacts with one with a negative opinion, her opinion flips with probability $(1-p)$ (compare this with the voter model where this occurs with probability 1). Similarly, for an agent with negative opinion, her opinion flips with probability $p$. So $p=1/2$ here is a case of unbiased flipping, which implies that flippings occur with 
probability $1/2$ whenever one interacts with an agent with the opposite opinion. In this model, 
one can formulate the mean field master equation as 
\begin{equation}
\frac{df_+}{dt} = f_{+}f_{-}p - f_+f_- (1-p).
\label{rand-flip}
\end{equation}
Interestingly, this coincides with Eq. (\ref{q=2}) such that one can interpret the 
$q=2$ model for any $p$ as a BMSBF. For $p=1/2$,  one gets $\frac{df_+}{dt}=0$ in Eq. (\ref{q=2}) which 
implies a conservation. This is expected as for $p=1/2$, we expect a voter model like behavior.

The solution of Eq. (\ref{q=2}) can easily be obtained as
\begin{equation}
f_+ = \frac{A e^{(2p-1)t}}{1+A e^{(2p-1)t}},
\label{sol-q2}
\end{equation}
where $A = \frac{f_+(0)}{1-f_+(0)}$.

Eq. (\ref{q=2}) shows that there are two fixed points $f_+ = f_+^* = 0,1$ for any $p$. Let $\delta$ be  defined as the infinitesimal deviation from a fixed point. Putting  $f_+ = f_+^* + \delta$, where $\delta$ is negative for $f_+^*=1$,  one gets up to linear order in $\delta$ (for small values of $|\delta|$), 
\begin{equation}
\frac{d\delta}{dt} = \delta (1-f_+^*)(2p-1) - \delta f_+^*(2p-1). 
\end{equation}
This leads to an  exponential time dependence  of $\delta$ as follows:
\begin{equation}
 |\delta| \propto e^{\pm (2p-1)t}\label{lyapunov}   
\end{equation}
where the $+$ ($-$) sign is for the fixed point $f_+^*=0$ ($f_+^* = 1$).
This shows that    $f_+^* =0$  is unstable (stable) for  $p >0.5$ ($p < 0.5$) as $\delta$ grows (decreases), and for $f_+^* =1$,
it is the opposite. This implies that whenever the initial configuration is biased towards the positive (negative) opinion, the final outcome would be a positive (negative) consensus for $p > 0.5$ ($p < 0.5)$. This will be reflected in the behaviour of the exit probability to be discussed later.

We also note that the so called Lyapunov exponents \cite{strogatz2018nonlinear} are same in magnitude for both the fixed points. Lyapunov exponents signify the (inverse of the) characteristic time scale with which infinitesimally close trajectories are separated in time $t$. From Eq. (\ref{lyapunov}), the Lyapunov exponents are $\pm (2p-1)$.

\subsubsection{$q=3$:}
In this case $n= 0, 1, 2, 3$. The $n=0$ and $n=3$ cases occur with 
probabilities $f_+^3$ and $(1-f_+)^3$ respectively. There can be six other cases
with $n=1$ or $2$ for which the probabilities in Eqs. (\ref{pq_plus}) and (\ref{pq_minus}) are obtained as the follows:\begin{align*}
&n=1: p_{3+} = \frac{p}{2-p} \text{ and }  p_{3-} = \frac{2(1-p)}{2-p}\\
&n=2: p_{3+} = \frac{2p}{p+1} \text{ and }  p_{3-} = \frac{1-p}{p+1}    
\end{align*}

The rate equation for $f_+$ is then given by
\begin{align}
\frac{df_+}{dt}=f_+^3(1-f_+)-(1-f_+)^3f_+ + \frac{2p}{p+1}3f_+^2(1-f_+)^2 + \frac{p}{2-p}
3f_+(1-f_+)^3 \nonumber\\
-\frac{1-p}{p+1}3f_+^3(1-f_+) - \frac{2(1-p)}{2-p}3f_+^2(1-f_+)^2,\label{q=3}
\end{align}
where the third and fourth terms on the right hand side are due to transition to a positive opinion while the last two terms are loss terms for $n=2$ and 1 respectively. Once again, the trivial fixed points are $f_+^* = 0,1$ and for $p=0.5$, any value is a fixed point. To perform a linear stability analysis, as we did for the previous case, we put $f_+=f_+^*+\delta$ in Eq. (\ref{q=3}) and ignoring higher order terms in $\delta$ we get
\begin{align}
\frac{d\delta}{dt}=3\delta {f_+^*}^2(1-f_+^*) - \delta {f_+^*}^3 - \delta(1-f_+^*)^3+3\delta f_+^*(1-f_+^*)^2\nonumber\\
+\frac{2p}{p+1}6\delta f_+^*(1-f_+^*)^2 - \frac{2p}{p+1}6\delta {f_+^*}^2(1-f_+^*)\nonumber\\
+\frac{p}{2-p}3\delta(1-f_+^*)^3 - \frac{p}{1-p}9\delta f_+^*(1-f_+^*)^2\nonumber\\
-\frac{1-p}{p+1}9\delta{f_+^*}^2(1-f_+^*) + \frac{1-p}{p+1}3\delta{f_+^*}^3\nonumber\\
-\frac{2(1-p)}{2-p}6\delta{f_+^*}(1-f_+^*)^2 + \frac{2(1-p)}{2-p}6\delta{f_+^*}^2(1-f_+^*)
\end{align}
Such that for $f_+^*=0$, 
\begin{equation}
\delta \propto e^{\frac{4p-2}{2-p}t}
\label{q=3delta1}
\end{equation}
and for $f_+^*=1$ (for which $\delta$ is negative), 
\begin{equation}
| \delta|  \propto e^{-\frac{4p-2}{1+p}t}.   
 \label{q=3delta2}
\end{equation}
Hence although the stability behavior of the fixed points are similar, the  Lyapunov exponents are dependent on the exact fixed points in contrast to the $q=2$ case.

\subsubsection{$q \geq 4$}

For larger values of $q$, taking $n=qf_+$ in Eqs. (\ref{pq_plus}) and (\ref{pq_minus}), 
 the transition rates $\omega$ between a positive $(+)$ state and a negative $(-)$ state are obtained as 
 \begin{align*}
    \omega_{-\rightarrow+}&=f_+^q+[1-f_+^q-(1-f_+)^q]\times\frac{pf_+}{(1-f_+)(1-p)+f_+p}\\[0.3cm]
    \omega_{+\rightarrow-}&=(1-f_+)^q+[1-f_+^q-(1-f_+)^q]\times\frac{(1-p)(1-f_+)}{(1-f_+)(1-p)+f_+p},
\end{align*}
where in the RHS, the first  terms correspond to conformity, i.e. when all the $q$ agents have the same opinion (positive and negative for the two equations respectively)  and the second terms include all other cases.   
Since  $$\frac{df_+}{dt}=-\omega_{+\rightarrow-}f_+(t)+\omega_{-\rightarrow+}f_-(t),$$
one therefore gets
\begin{equation}
    \frac{df_+}{dt}=f_+^q(1-f_+)-(1-f_+)^qf_++(1-f_+)f_+\frac{\{1-f_+^q-(1-f_+)^q\}(2p-1)}{(1-p)(1-f_+)+pf_+}.
    \label{MF}
\end{equation} It maybe noted that there are two fixed points $f_+^* =0,1$ for all values of $p,q$ and also a third, which is obtained after numerically solving the equation. The third fixed point in general, depends on both $p$ and $q$.  Also, if one puts $f_+=f_- = 0.5$ and $p=0.5$ in the above equation, one gets $\frac{df_+}{dt} = 0$,
which implies that for any $q$ this is the third fixed point. 

For $q \to \infty$ limit, the Eq. (\ref{MF}) becomes
\begin{equation}
    \frac{df_+}{dt}=(1-f_+)f_+ \frac{(2p-1)}{(1-p)(1-f_+)+pf_+}\label{MF2}
\end{equation}
such that the fixed points are again simply $f_+^* = 0,1$ and for $p =0.5$, 
all points are fixed points.



For large $q$ one can take Eq. (\ref{MF2}) and perform a linear stability analysis by substituting $f_+=f_+^*+\delta$ to get \begin{equation}
    \frac{d\delta}{dt}=
    \delta\frac{[1-2f_+^*](2p-1)}{(1-p)(1-f_+^*)+pf_+^*} + \mathcal{O}(\delta^2)
\end{equation} 
Thus, for $f_+^*=0$ we get
\begin{equation}
    \frac{d\delta}{dt}=\frac{2p-1}{1-p}\delta.\label{p.gt.0.5}
\end{equation} 
This indicates growth of $\delta$ for $p>0.5$. Similarly for $f_+^*=1$ we get,
\begin{equation}
    \frac{d\delta}{dt}=-\frac{2p-1}{p}\delta,\label{p.lt.0.5}
\end{equation} which indicates growth of $f_+(t)$ for $p<0.5$. Since $\delta$ can not exceed $1$ or be less than $-1$ for the two regions where we find growth (decay) of $f_+(t)$, we need vanishing contribution from $\delta$, i.e., $f_+=1+\delta$, where $\delta<0$ (say for $f_+^*=1$). Then using Eq. (\ref{p.gt.0.5}) we get, \begin{equation}
    f_+=1-|\delta|=1-|\delta_0|e^{-\big(\frac{2p-1}{p}\big)t}\label{delta2}
\end{equation} Similarly, for $f_+^*=0$, say $f_+=\delta$ and $\delta=\delta_0e^{-\big(\frac{2p-1}{1-p}\big)t},$ using Eq. (\ref{p.lt.0.5}). This vanishes for $p<0.5$. So in this region we have \begin{equation}
    f_+=\delta_0e^{\big(\frac{2p-1}{1-p}\big)t}\label{delta1}.
\end{equation}

Although here we considered $q \geq 4$ as the calculations for $q=2,3$ are done considering all possible values of $n$, one can still put $q=2$ or 3 in Eq. (\ref{MF}) which leads to    a third fixed point as mentioned earlier (e.g, for $q=2$, the third fixed point is at $(1-p)$ for $p \neq 0.5$).  However, this is not true according to the calculations done without assuming $n=qf_+$. Apparently the third fixed point is an artefact of this assumption. We will get back to this point later.

\subsection{Monte Carlo simulations}\label{mcs}

 We also use Monte Carlo simulations, and compare these results with that found using mean-field approach. The simulation begins with $x$ fraction of agents with positive opinion, i.e., $f_+(t=0)=x$. For our scheme of simulation we use random asynchronous update, which means that at each Monte Carlo (MC) time step $L$ agents are randomly chosen and are instantly updated according to the defined dynamical rule, where $L$ is total number of agents in the system. Let us summarise the dynamical rule for our model as defined in section \ref{model}: \begin{enumerate}
    \item Randomly select an agent $i$ from $1$ to $L$.
    \item Randomly select $q$ other agents such that none of these are the same as $i$. This was done using Fisher-Yates shuffle algorithm \cite{fisher1963statistical}. These $q$ agents are selected without repetition, and termed as the $q$-panel with which agent $i$ interacts.
    \item If all of these $q$ agents have the same opinion, i.e., the $q$-panel is unanimous, then agent $i$ takes this opinion -- a situation defined as conformity.
    \item Otherwise, if the $q$-panel is non-unanimous, then we count the number of agents with positive opinion $n$ in this panel, and then let agent $i$ take the positive opinion with probability $p_{q+}$ and the negative opinion with probability $p_{q-}$. The expressions for $p_{q+}$ and $p_{q-}$ are given by Eqs. (\ref{pq_plus}) and (\ref{pq_minus}).
    \item Steps 1 to 4 are repeated $L$ times. This constitutes one MC time step.
    \item The simulations are then continued until a maximum number of MC steps, or until a global consensus is reached, i.e., each of $L$ agents in the system has the same opinion.
    \item Finally the results are averaged over several initial configurations.
\end{enumerate}

\begin{figure}[h!]
    \centering
    \includegraphics[width=\textwidth]{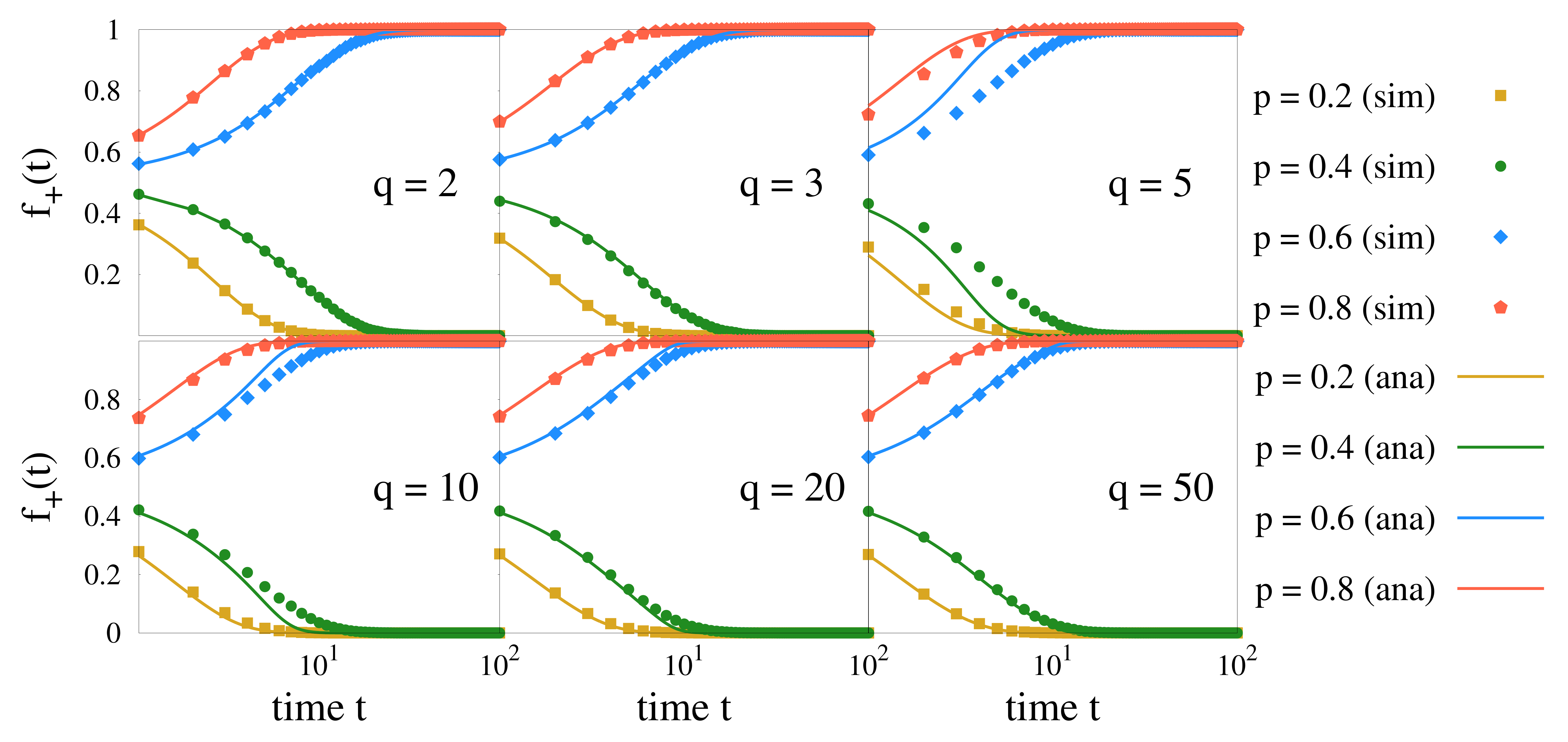}
    \caption{Variation of fraction $f_+$ of agents with positive opinion as a function of time $t$ for different values of $p$ \& $q$, and $x=0.51$. Simulated (sim) results are shown by solid circles, and analytical (ana) results are shown by solid lines. Simulations are performed using $L = 1024$ averaging over $100$ configurations. The agreement between simulated and analytical results are excellent for $q=2$ and $3$. For cases with $q\geq 4$ the agreement becomes more reasonable as $q$ increases.}
    \label{sim_vs_ana1}
\end{figure}

 First we calculate the fraction $f_+(t)$ of up spins as a function of time $t$. The results are shown in Figure \ref{sim_vs_ana1}.  To compare the numerically simulated results with our analytical expressions, we use Eq. (\ref{sol-q2}) for the case $q=2$, as it gives an exact solution for $f_+(t)$. However, we use Euler's method to solve the differential equations given by Eq. (\ref{q=3}) for $q=3$ and Eq. (\ref{MF}) for $q\geq 4$. We observe that for $q=2$ and $3$, numerical and analytical results match in an excellent manner. For finite $q$ values $\geq 4$, there is a discrepancy which vanishes as $q$ is made larger. As $q$ increases beyond $4$ the agreement between numerical and analytical results becomes approximate. Once again in the large $q$ limit (e.g. $q=50$) the agreement becomes excellent.  The reason for this could be understood as follows.

\begin{figure}[h!]
    \centering
    \includegraphics[width=\textwidth]{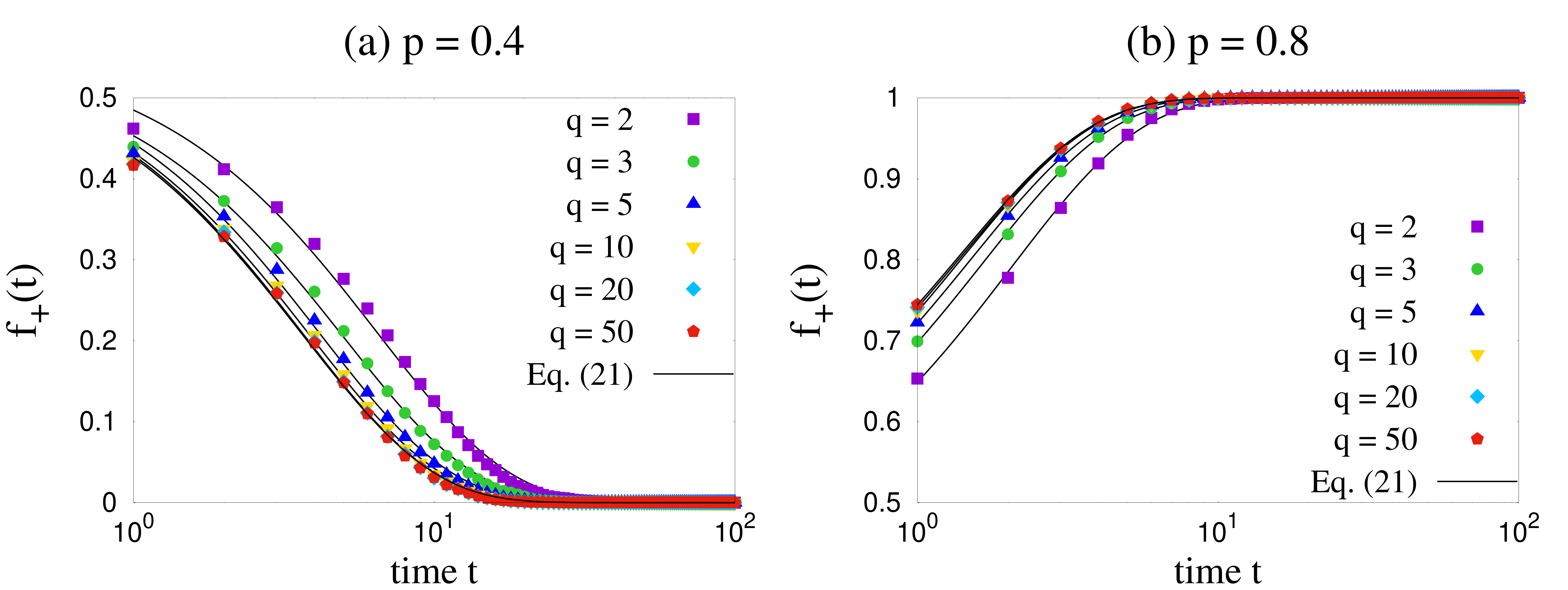}
    \caption{Fraction $f_+$ of agents with positive opinion as a function of time $t$ for $x=0.51$ for several values of $q$ and two typical values of $p$, viz. (a) $p=0.4$ and (b) $0.8$. Simulations are done for $L=1024$ on a complete graph. As $q$ increases, $f_+$ becomes $q$ independent. The black curves are data fittings done using Eq. (xx).}
    \label{frac_q_indep}
\end{figure}

 In section \ref{mfa} for $q\geq 4$ we had assumed that within the $q$ panel there are $f_+$ fraction of agents with positive opinion, which makes $n=qf_+$. This argument holds well from a theoretical point of view. However, in the course of simulations $q$ and correspondingly $n$ values have to be integers, but $f_+$ is a fraction such that $0<f_+<1$. So the validity of $n=qf_+$ becomes very approximate for low $q$, which clearly is less problematic for higher values of $q$. Thus we observe excellent agreement between numerical and analytical results for higher values of $q$.

 Overall the general behavior of $f_+(t)$ for different values of $p$ as observed from Figure \ref{sim_vs_ana1} is trivial. For $p<0.5$ the fraction of agents with positive opinion becomes $0$, and for $p>0.5$ this fraction becomes $1$. This implies that when the influential power of agents with positive opinion is less than that of the agents with negative opinion, the system reaches a negative consensus and vice-versa for the case when the influential power of agents with positive opinion is greater than that of the agents with negative opinion. To understand this further we studied the exit probability $E(x)$ for several values of $p$ in our model.

\subsection{Exit Probability}\label{exit}

\begin{figure}[h!]
    \centering
    \includegraphics[width=\textwidth]{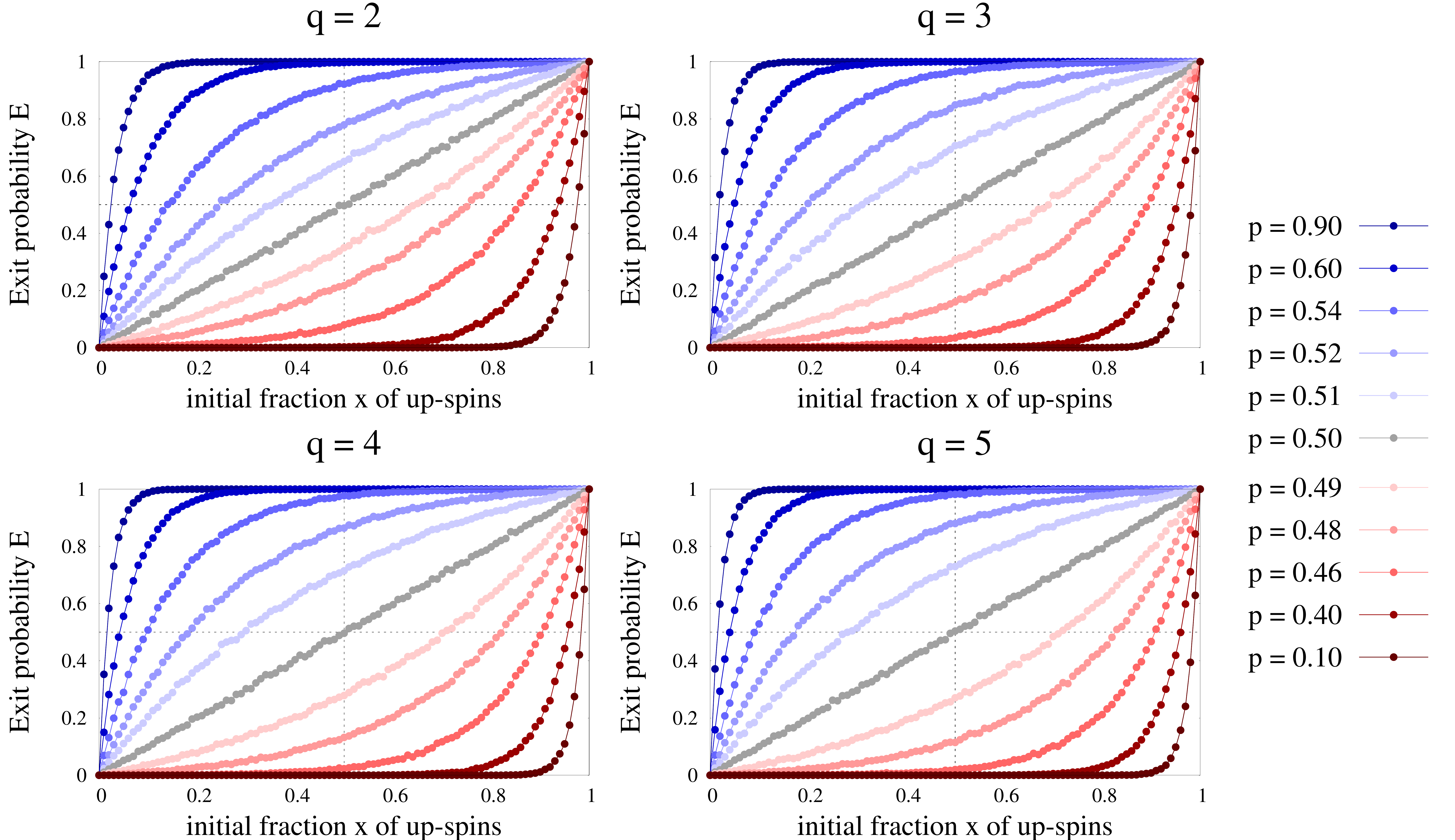}
    \caption{Exit probability $E(x)$ as a function of initial fraction $x$ of agents with positive opinion for several values of $p$ and 4 values of $q$. Simulation were done for $L = 64$ on a complete graph. The results are qualitatively similar across the values of $q$ shown here.}
    \label{exit_combined}
\end{figure}

\begin{figure}[h!]
    \centering
    \includegraphics[width=\linewidth]{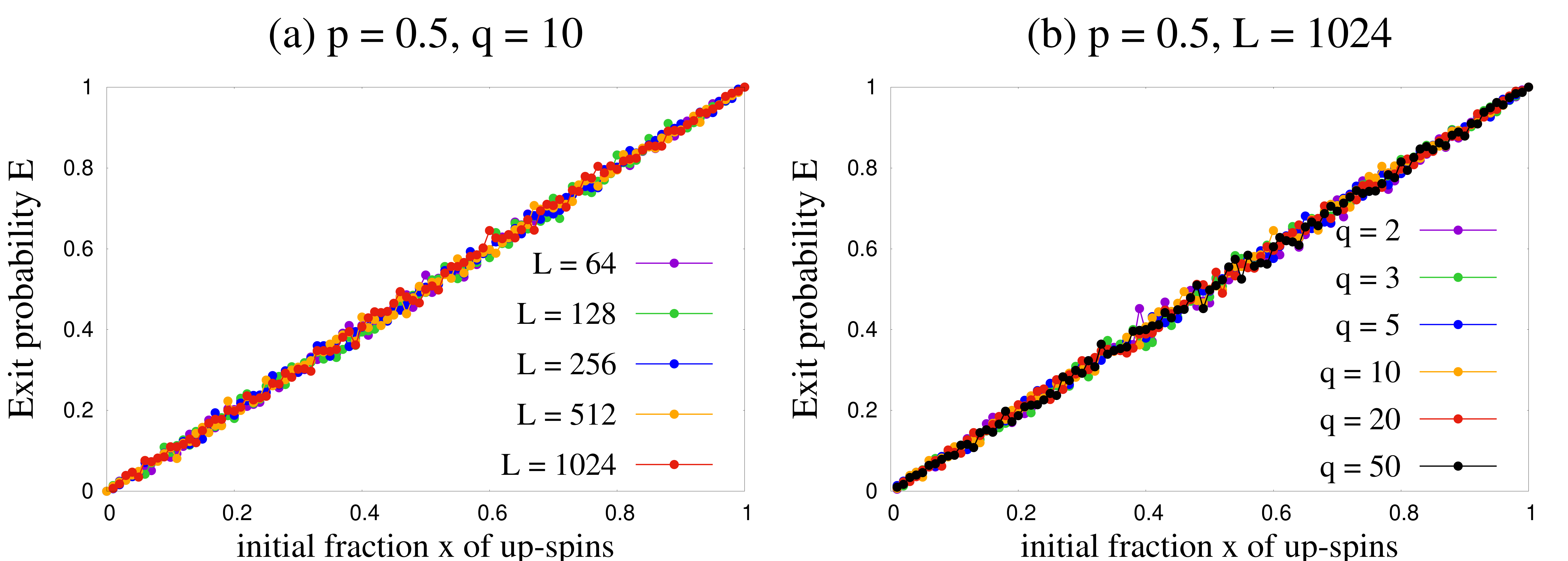}
    \caption{Variation of exit probability $E(x)$ as a function of initial fraction $x$ of up spins, for (a) several system sizes with $p=0.5$, $q=10$ and for (b) various values of $q$ with $p=0.5$, $L=1024$. It seems that $E(x)$ maintains its linear behaviour even in the thermodynamic limit and in large $q$ limit.}
    \label{exit_L_q_dep}
\end{figure}

Our results for exit probability are shown in Figure \ref{exit_combined}.  It seems that the results are independent of $q$ when we study $E(x)$ for lower values of $q$. Before discussing further on dependence of exit probability results on $q$, let us first focus on the point $p=0.5$. From Figure \ref{exit_combined} we can see that $E(x)=x$ for $p=0.5$ for all the values of $q\leq 5$. Figure \ref{exit_L_q_dep} shows the results for $p=0.5$ several values of $L$ and $q$, and they are consistent with $E(x)=x$. 
 This is expected from our discussions in sec \ref{model} where we argued that for $p=0.5$, the model becomes equivalent to a voter model. Such a behaviour, however, is not apparent from the mean field theory except for very large $q$ values when the first two terms in Eq. (\ref{MF}) could be ignored.

\begin{figure}[h!]
    \centering
    \includegraphics[width=\textwidth]{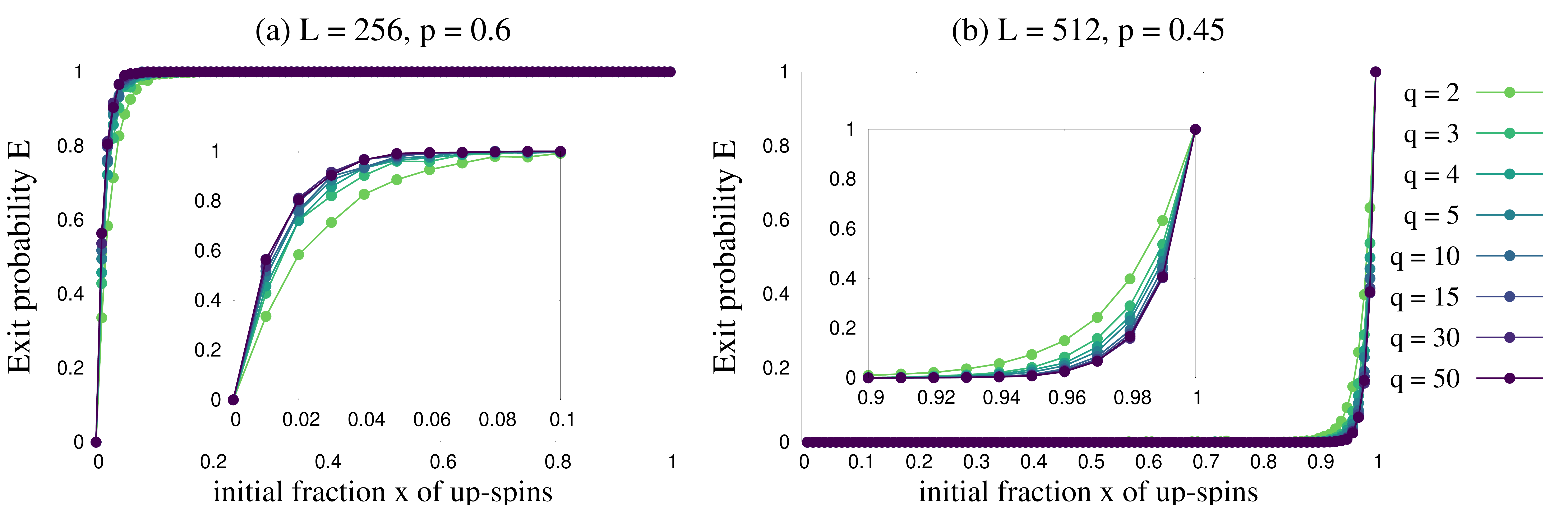}
    \caption{Exit probability $E(x)$ as a function of initial fraction $x$ of agents with positive opinion for several values of $q$ from $2$ to $50$ and for (a) $L=256$, $p=0.6$ and (b) $L=512$, $p=0.45$. The results converge as $q$ grows larger.}
    \label{exit_q_dep}
\end{figure}

 Next we focus on the dependence of exit probability results on $q$ for general $p$ values. As already mentioned, for lower values of $q$ the exit probability curves are qualitatively $q$-independent, as seen from Figure \ref{exit_combined}. To dig into this further, we numerically obtain $E(x)$ versus $x$ curves up to $q=50$, and show their comparison with lower values of $q$ in Figure \ref{exit_q_dep}.  The exit probability curves actually converge as $q$ takes larger values. This means that the steady states  in our model do not depend on $q$ as the value of $q$ increases. This was also confirmed when we studied the trajectories for fraction $f_+$ of agents with positive opinion (Figure \ref{frac_q_indep}).

\begin{figure}
    \centering
    \includegraphics[width=\textwidth]{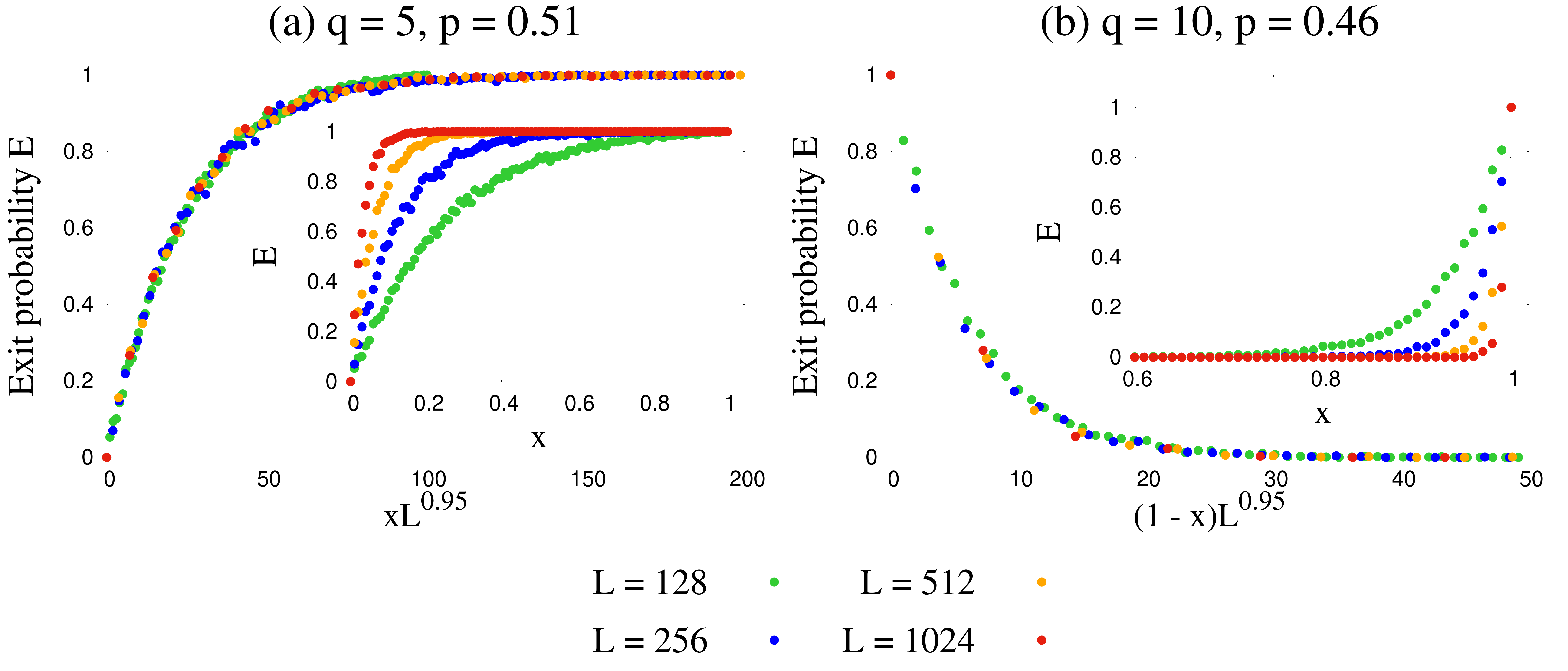}
    \caption{Data collapse of exit probability $E(x)$ for several values of system size $L$ from $128$ to $1024$ for (a) $q=5$, $p=0.51$ and (b) $q=10$, $p=0.46$. For $p>0.5$ the scaling argument is $E(x)\sim xL^{0.95}$ and it is $E(x)\sim (1-x)L^{0.95}$ for $p<0.5$. Insets show the unscaled data. It is evident that in the thermodynamic limit $L\rightarrow\infty$ the exit probability would become a step functions at $x=0$ for $p>0.5$ and at $x=1$ for $p<0.5$.}
    \label{exit_L_dep}
\end{figure}

 But how do our results depend on the system size $L$? To investigate the finite size effect of exit probability we simulated our model for several system sizes from $L=64$ to $1024$ and have shown the results in Figure \ref{exit_L_dep}.  It is clear that in the thermodynamic limit $L\rightarrow\infty$ the exit probability exhibits a step function like behavior even for a minor deviation from $p=0.5$. We can see from Figure \ref{exit_L_dep} that for $p>0.5$ the exit probability would show a step function at $x=0$, and for $p<0.5$ it would show a step function at $x=1$ in thermodynamic limit. The implication of this observation is very critical from the perspective of opinion dynamics in human societies. It seems that even if in the beginning we have a very small fraction $x$ of agents with positive opinion, the system could still reach a positive consensus given the influential power $p$ of agents with positive opinion is slightly higher than that of agents with negative opinion. Quite similarly, if initially we have a very large fraction $x$ of agents with positive opinion, the system could still reach a negative consensus given the influential power $p$ of agents with positive opinion is slightly smaller than that of agents with negative opinion. The parameter $p$, the influential power of agents with positive opinion, thus introduces a broken symmetry in the dynamics, irrespective of the value of $q$. The obtained data for exit probability $E(x)$ was collapsed using the following scaling form: \begin{align}
    E(x)&\sim xL^{\nu}\hspace{0.2cm}\text{for }p>0.5\nonumber\\
    &\sim (1-x)L^{\nu}\hspace{0.2cm}\text{for }p<0.5
\end{align} where we found a universal value of $\nu \simeq 0.95$ for any $q$. The collapsed data was found to fit well according to the functional form $1-a\exp{(-bx)}$ for $p>0.5$ and $a\exp{(-bx)}$ for $p<0.5$, where $a$ is a constant and the parameter $b$ becomes $q$-independent as $q$ increases. This implies that $b$ is like a scale governing the approach to unity for $p>0.5$ or to zero for $p<0.5$ for the scaled exit probability.

\begin{figure}
    \centering
    \includegraphics[width=\textwidth]{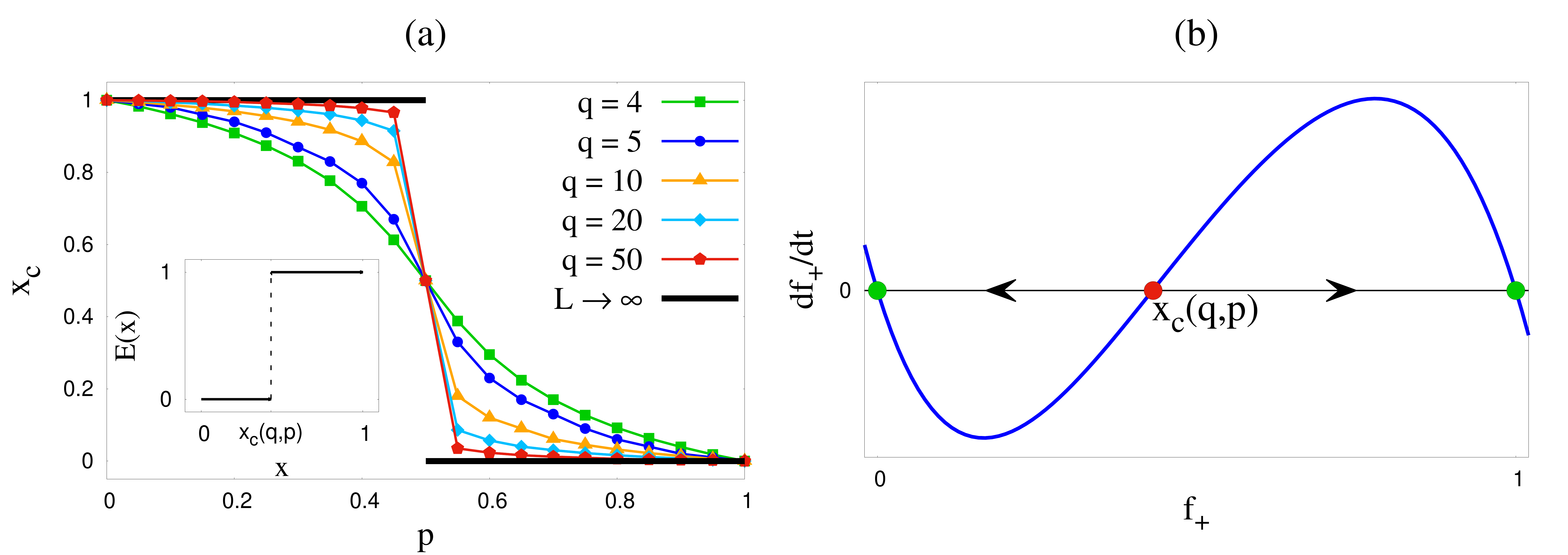}
    \caption{(a) Variation of $x_c$ as a function of $p$ for several values of $q$ as obtained by mean-field theory. The solid line in black denotes the results from Monte Carlo simulation for $L\rightarrow\infty$. In large $q$ limit, the mean-field results converge to Monte Carlo results. \textit{Inset} shows the behavior of exit probability as expected from the mean-field estimations. (b) Generalised phase portrait for our model, where the green circles indicate stable fixed points at $0$ \& $1$, and the red circle denotes the unstable fix point at $x_c$. The values of $x_c$ depend on $q$ and $p$. The arrows indicate the directions of flow, such that for $f_+(t=0)<x_c$, we would have $f_+ (t\rightarrow\infty)=0$ and similarly $f_+ (t\rightarrow\infty)=1$ for $f_+(t=0)>x_c$.}
    \label{euler_results}
\end{figure}

 This observation could however not be made from the mean-field results. If we define $x_c$ as the cut-off value of initial fraction $x$ of agents with positive opinion below which the system reaches a negative consensus, i.e., exit probability shows a step function (in the thermodynamic limit), then according to Monte Carlo results $x_c=1$ for $p<0.5$ and $x_c=0$ for $p>0.5$. However, the mean-field results show the existence of a non-trivial $x_c$ for each $q$, as summarised in Figure \ref{euler_results}(a). So $x_c$ is basically the unstable fixed point, as shown in Figure \ref{euler_results}(b). $x_c$ was estimated by numerically solving Eq. (\ref{MF}) and finding $x$ below which the saturation value of $f_+$ is $0$.  Although as $q$ increases, analytically obtained values of $x_c$ approaches that obtained by numerical simulation. This once again confirms that in the large $q$ limit mean-field results converge to Monte Carlo results.

 Interestingly in the mean field theory, for any $q$,  $x_c=0.5$ for $p=0.5$ (this is consistent with the discussions at the end of sec \ref{mfa} for any finite $q$, or for $q\rightarrow\infty$). This implies that for $p=0.5$, exit probability should show a step function at $x=0.5$, according to mean-field theory. However for infinite $q$, all points are fixed points for $p=0.5$ which will give a linear exit probability.  In the Monte Carlo simulation, as shown in Figures \ref{exit_combined} and \ref{exit_L_q_dep}, exit probability indeed shows a linear behavior with $x$ at $p=0.5$ for several values of $q\leq 50$ and also for several system sizes. 
 A minor deviation from $p=0.5$ would change this linear behavior to a step function like behavior in the thermodynamic limit as shown in Figure \ref{exit_q_dep}, which then agrees with mean-field result in the large $q$ limit.

\subsection{Dynamics}

\subsubsection{Relaxation behavior}
\label{relax}


So far, the main result is that the $p=0.5$ point separates the two regions of consensus with positive opinion (for $p> 0.5$) and negative opinion (for $p < 0.5$). This corresponds to the 2 fixed points $f_+^*$ discussed in section \ref{mfa}. 
An interesting question is how, from arbitrary initial configurations, the system evolves towards either of the fixed points, i.e., the relaxation behaviour and the associated time scale, if any. In case of an exponential growth/decay of the relevant quantities, it is possible to define such a time scale (note that this is different from the exact time to reach the fixed point). 

We found that the qualitative behavior of exit probability as well as the value of the
exponent $\nu$ are independent of the exact value of $q$.  Here, we report  how the dynamics are affected by the value of $q$.
 Eq. (\ref{sol-q2}) shows that when $q=2$, for any initial value, at large but finite times, the behavior of $f_+(t)$ is either $1 - \alpha e^{-\beta t}$  or $\alpha^\prime  e^{-\beta^\prime t}$ for $p < 0.5$ with $\beta = 2p-1$. 

We conjecture that for any $q$,  $f_+$  will have  the form 
\begin{align}
    f_+(t)&=\alpha e^{-\beta t},\hspace{0.33cm}\text{for }p<0.5\nonumber\\
    &=1-\alpha^\prime e^{-\beta^\prime t},\hspace{0.33cm}\text{for }p>0.5.
    \label{falphabeta}
    \end{align} 
    In general $\beta$ and $\beta^\prime$, functions of $p$, can be different as found for $q=3$ (Eqs. \ref{q=3delta1}, \ref{q=3delta2}). The coefficients $\alpha, ~ \alpha^\prime $ are trivially related to the initial values.
    
\begin{figure}
    \centering
    \includegraphics[width=\linewidth]{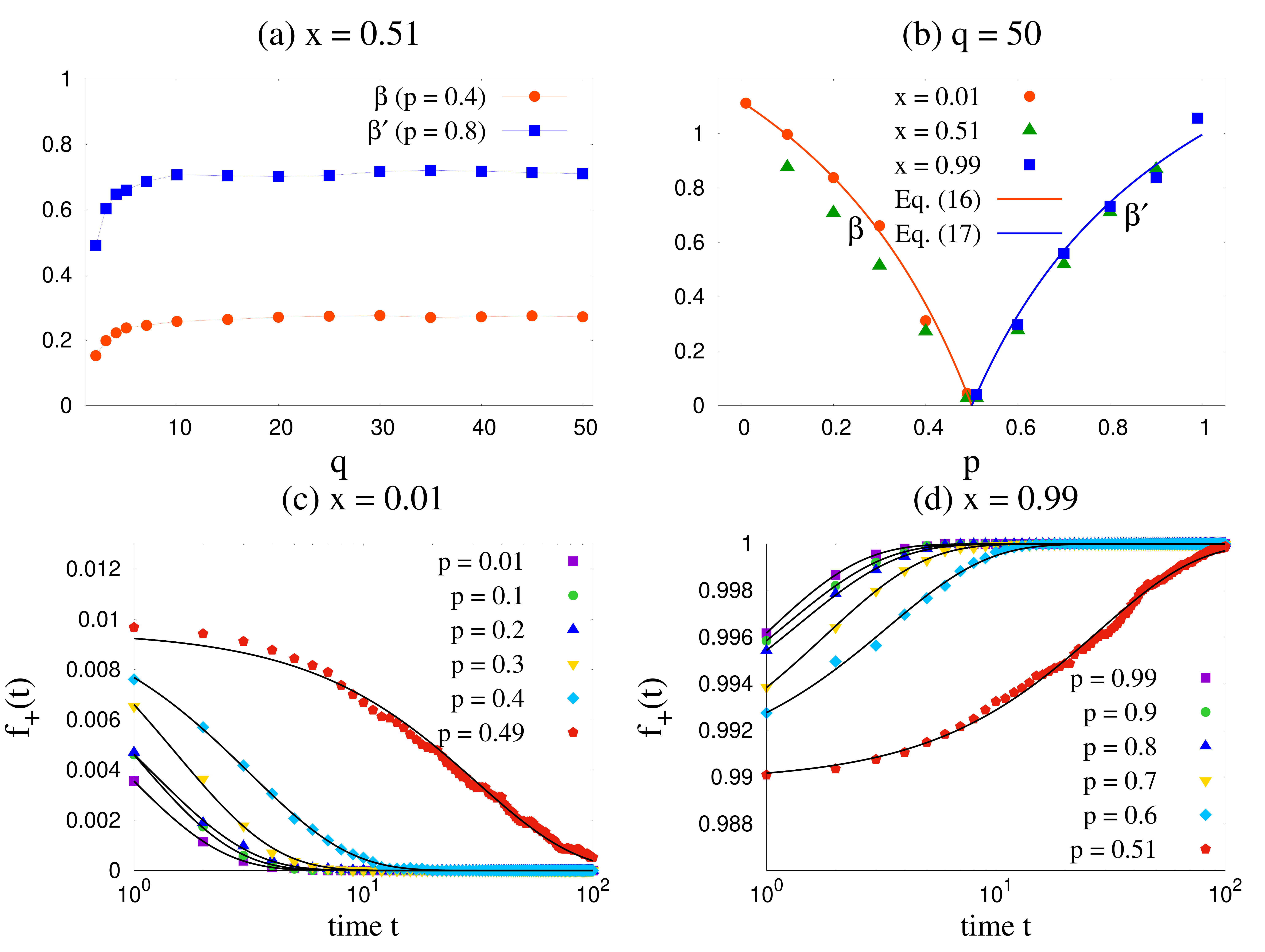}
    \caption{(a) shows the obtained values of $\beta$ and $\beta'$ as a function of $q$ for 2 typical values of $p$ keeping $x$ fixed. In (b) we show the obtained values of $\beta$ and $\beta'$ as a function of $p$. The data is fitted according to Eqs. (16) and (17). (c) and (d) shows the fitting of the $f_+(t)$ curves for $q=50$ according to Eq. (15) for 2 typical values of $x$, viz. $x = 0.01$ in the region $p<0.5$ and $x = 0.99$ in the region $p>0.5$ respectively.}
    \label{exp.fit}
\end{figure}

In Figure \ref{frac_q_indep} we fit the time dependence of $f_+$  in the above form for two typical values of $p$, and for several values of $q$ from $2$ to $50$.  The results for $\beta, \beta^\prime$ are different for the two values of $p$ as expected but become independent of $q$ for large values of $q$, as shown in Figure \ref{exp.fit}. For small $q$ values there is an increase with  $q$.  

In order to obtain the  dependence of $\beta, \beta^\prime $ on $p$ from the mean field theory for large $q$, we take note from 
Eqs. (\ref{delta2}) and (\ref{delta1}) the variation of $\delta$ (which is linearly related to $f_+$) with $p$.  We argue that since $\delta$ cannot increase 
indefinitely, it is advisable to extract the values of the parameters from their
vanishing feature. For the region $p < 0.5$, $\delta$ goes to zero in
Eq. (\ref{p.lt.0.5}) and for $p > 0.5$, $\delta$ goes to zero in Eq. (\ref{p.lt.0.5}).
Now the expressions for $f_+ = f_+^* + \delta $ with $ f_+^* = 0, 1$ 
are in the form of eqs \ref{falphabeta} with the values of $\beta, 
\beta^\prime$ given by 
\begin{align}
   \beta(p)&\sim \frac{1-2p}{1-p}\hspace{0.33cm}\text{for }p<0.5\\
   \beta^\prime(p)&\sim \frac{2p-1}{p}\hspace{0.33cm}\text{for }p>0.5
   \label{betas}
\end{align}

In Fig \ref{exp.fit}, the results for $f_+$ for a large value of $q=50$ and different $p$ values are shown from which the numerical estimates of $\beta, \beta^\prime$ are made. The comparison with the theoretical estimates shows very good consistency. To extract the values of the parameters and compare them with the theoretical ones, 
we used  $x = 0.01$ and $x=0.99$ as the initial values of $f_+$ in the two regions $p < 0.5$ and $p > 0.5$ respectively as the linear stability analysis is valid for  small $\delta$ deviating from the fixed points zero and 1. We have found that for other values of $x$ also, the $p$ dependence of $\beta, \beta^\prime$ are similar, apart from some trivial multiplicative factors. One can conclude from this that there is a timescale which diverges as $p \to 0.5$ from either side. Both the  timescales are inversely proportional to $\beta, \beta^\prime$ and therefore $\propto (2p-1)^{-1}$. The point $p=0.5$ can therefore be interpreted as a dynamical critical point manifesting critical slowing down.


\begin{figure}
    \centering
    \includegraphics[width=\linewidth]{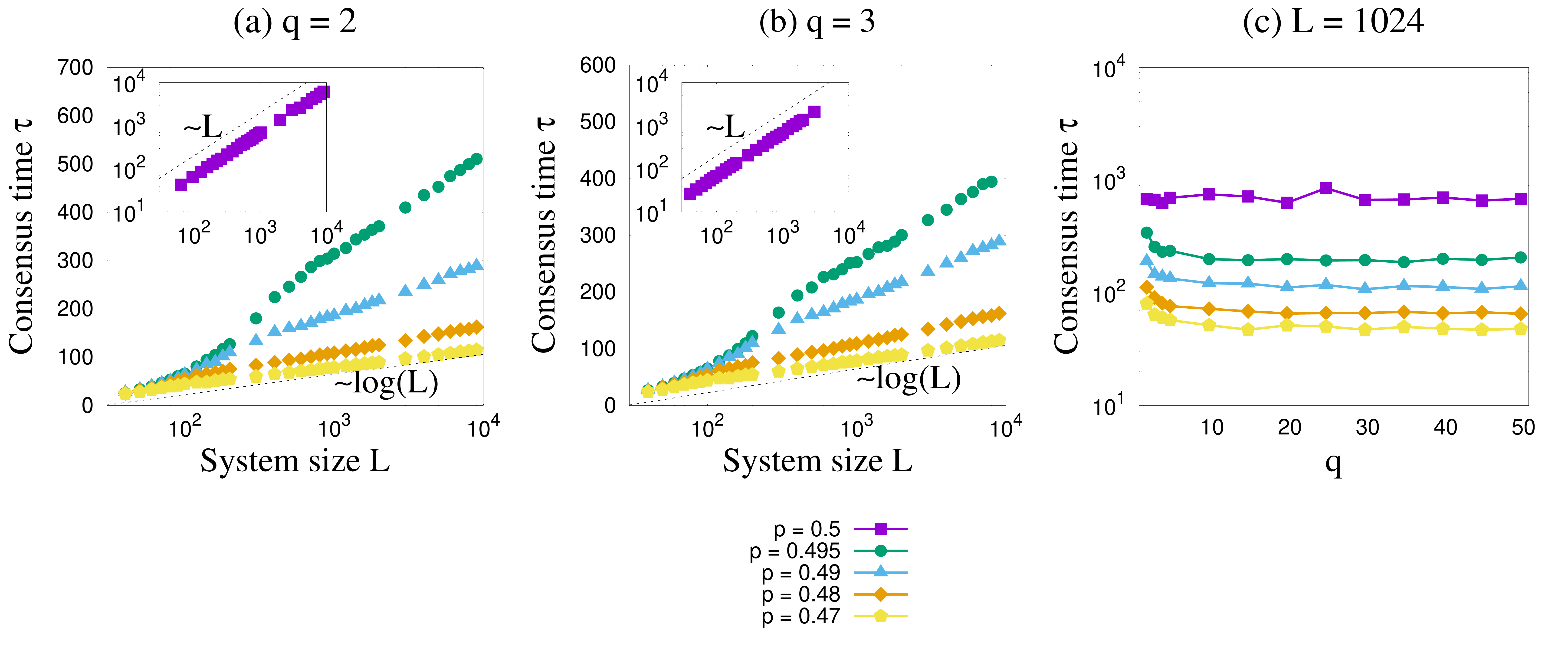}
    \caption{Consensus time $\tau$ as a function of system size $L$ for several values of $p$ for (a) $q=2$ and (b) $q=3$. Insets show the data for $p=0.5$. The simulations were done for $x=0.5$. We can see that $\tau\sim L$ for $p=0.5$, but as $p$ deviates from $0.5$ the variation takes a logarithmic form. In (c) we show the variation of $\tau$ as function of $q$ for system size $L=1024$. $\tau$ decreases for lower values of $q$, however it does not exhibit a systematic dependence as $q$ is made larger.}
    \label{con.time}
\end{figure}

\subsubsection{Consensus times: dependence on system size}
\label{constime}

From the Monte Carlo simulations, one can estimate the time to reach the consensus states as a function of the system size. For the unbiased case,  the dependence is a linear relation while for $p \neq 0.5$ the results  indicate a logarithmic variation, as also shown in \cite{krapivsky2021divergence}. The linear relation is also found in the mean field voter model. The results clearly show that the  dynamics are much faster for any value of $p$ different from 0.5. Figure \ref{con.time} shows that the variation of the times are nearly independent of $q$ as $q$ is made larger consistent with the other results obtained.


\section{Conclusion}\label{conclu}

In this paper, we have studied a dynamical model of opinion formation where the opinion of an individual is determined on the basis of the opinions of other $q$ number of agents. These $q$ agents are basically the social connections with which the individual has an interaction. A parameter $p$ determines the influential power of the agents with positive opinion. So $p$ acts as a bias in the system; for $p=0.5$, the model is identical to a mean field voter model. We analysed the dynamics of the system in terms of the fraction of agents with positive/negative opinion and also the steady states. 

We obtain three regions, $p < 0.5$, $p=0.5$ and $p > 0.5$ that determine the fate of the system.  Interestingly, the equilibrium results for the ensemble averaged quantities, are independent of $q$ in the thermodynamic limit. It implies that the size of the social connections influencing an agent is irrelevant and only the bias $p$ matters as found from the simulation results. 
However, at small $q$ values, the results are quantitatively $q$ dependent, for example, the relaxation timescale (see section \ref{relax}). We argue that as $q$ increases, the fluctuations in the opinions in the $q$ plaquette becomes less effective and as a result one gets $q$ independent behavior for large $q$. This is analogous to mean field theory being valid at higher dimensions in general.


The mean field results for $q=2,3$, derived with all possible composition of the $q$ plaquette show very good agreement with the simulation results. In these cases, as well as in the simulations, there are only two fixed points for $p \neq 0.5$. On the other hand, the mean field rate equations derived for higher values of $q$ are formulated assuming an average  number of  $n=qf_+$ agents  with opinion $+1$ in the $q$ plaquette. So for all the cases where there is no unanimity, a single configuration is considered with this value of $n$. This assumption implies that the distribution of opinions in the $q$ plaquette, which is  a subset of the whole system, 
is taken to be identical to the bulk and fluctuations are ignored. Comparison with numerical simulations indicates that the existence of the third fixed point in the mean field theory results from this assumption. However, we found by numerically solving the equations that this third fixed point is an unstable one so except for the case when one starts from exactly at the fixed point, the final states are consensus states with either all positive/all negative opinions. If one starts with a value of $f_+$ above (below) this fixed point, the all positive (negative) consensus state is reached. Therefore the step function for exit probability, according to mean field equations, occurs at a $p,q$ dependent value. Of course for $q$ very large, the mean field theory also
shows the existence of two fixed points; this happens as neglecting fluctuations in the
$q$ plaquette does not affect the results anymore. 

The outcome of the study is quite simple but not obvious: even an infinitesimal initial bias towards the positive opinion will lead to a consensus state with all agents having positive opinion when the bias $p$ is greater than $0.5$. For $p < 0.5$, similarly, the all negative consensus state is reached for an initially biased (however small) state towards negative opinions. So the effect of minority spreading can occur here and these results are independent of $q$ in the thermodynamic limit. The exponent  obtained from the data collapse of  the exit probability is shown to be universal and very close to unity.

While the equilibrium features of the model are independent of $q$, the dynamical behavior do show $q$ dependence, at least for small $q$. We have made a linear stability analysis for $q=2,3$ and $q \to \infty$ to show that the corresponding dynamical behavior are different. However, again we find that as $q$ is made larger, the exponents are independent of $q$. One interesting observation is that one can identify $p=0.5$, the mean field voter model point, as a dynamical critical point as  critical slowing down occurs close to it.  The corresponding timescale diverges with an universal exponent equal to unity according to the mean field theory and supported by numerical estimates. The system size dependence of the consensus time $\tau$ is again  identical for all $q$ and shows a logarithmic dependence for $p \neq 0.5$. For $p = 0.5$, it is linear. Hence this provides further proof that the system behaves very differently 
as $p$ deviates from 0.5. 

The $q$-voter model with weighted influence presented in this study has significant potential applications in various domains where consensus building and opinion dynamics play a crucial role, such as strategic management, information retrieval, and human resources management. By capturing the effects of social influence and biased opinion spread, this model can be used to optimize decision-making processes, improve network management, and analyze information dissemination in intelligent systems. Furthermore, its relevance to multi-agent systems and knowledge discovery makes it a valuable tool for developing and enhancing systems in fields such as finance, marketing, and crisis management.

In conclusion, we presented a model, where one of the opinion holds an ``edge" making the agents with that opinion more influential. Out of the two parameters used, $p$ happens to determine 
the qualitative behavior entirely. Quantitative results become $q$ independent as $q$ is made larger.

\section*{Acknowledgements}

PS thanks the hospitality provided by Wrocław University of
Science and Technology. Katarzyna Sznajd-Weron is thanked for a critical reading of the manuscript.

\section*{Funding}

PM did not receive financial support for this research. PS received financial support from (i) SERB, Government of India (grant no. MTR/2020/000356) and (ii) CSIR, Government of India (file no. 03/1495/23/EMR-II). 

\section*{Disclosure Statement}

The authors report that there are no competing interests to declare.

\section*{Data availability statement}

All the simulated data has been presented in manuscript in the form of plots. The code used for simulation could be made available upon reasonable request.

\bibliographystyle{elsarticle-num} 
\bibliography{mbibliography}

\end{document}